\documentclass[twocolumn,showpacs,preprintnumbers,amsmath,amssymb]{revtex4}  

\usepackage{epsfig,psfrag}  
\usepackage{dcolumn}
\usepackage{bm}
\usepackage{graphicx}  
\usepackage{color}  
  
\begin{document}  
  
  
\title{Full counting statistics for the Kondo dot.}  
  
\author{A.~Komnik$^1$ and  A.O.~Gogolin$^2$}  
  
\affiliation{${}^1$~Physikalisches Institut,  
Albert-Ludwigs-Universit\"at,  
  D-79104 Freiburg, Germany \\  
${}^{2}$ Department of Mathematics, Imperial College London,  
180 Queen's Gate, London SW7 2AZ, United Kingdom   
}  
  
\date{\today}  
  
\begin{abstract}  
The generating function for the cumulants of charge  
current distribution is calculated for two   
generalised Majorana resonant   
level models: the Kondo dot at the Toulouse   
point and the resonant level embedded in a Luttinger liquid    
with the interaction parameter $g=1/2$. 
We find that the low--temperature non-equilibrium transport in the Kondo case
occurs via tunnelling of physical electrons as well as by coherent transmission
of electron pairs.
We calculate the third cumulant (`skewness') explicitly and analyse it  
for different couplings, temperatures, and magnetic fields.  
For the $g=1/2$ set-up the statistics simplifies and is given by a  
modified version of the Levitov--Lesovik formula.  
\end{abstract}  
  
  
\maketitle  
  
Since Schottky's realisation that the shot noise in a  
conductor contains invaluable information about the physical  
properties of the charge carriers, the question about the noise  
spectra of different circuits became as important as the knowledge  
of their current--voltage characteristics \cite{schottky}.   
The noise constitutes the second moment of the  
current distribution function (which is the probability of  
measuring a given value of the current) and is supposed to contain information
about the charge of current carrying excitations at 
weak transmission (reflection). That is indeed the case for $S-S$ and $S-N$
junctions \cite{muz,cuevas,jehl} but not been proven for a generic
interacting model. As has been pointed out in \cite{reznikov},
the third cumulant also contains valuable information about the
charge of the current carriers. Therefore it is 
natural to investigate the full current distribution function.   
This was an academic question for a very long time as even   
the measurement of the second cumulant remained on  
the frontier of experimental physics.   
Only after the work by Reulet and coworkers the  
measurement of the third cumulant became possible \cite{reulet}.   
Inspired by this remarkable achievement, the full current  
distribution function (more often referred to  
as `full counting statistics' or FCS) has been   
theoretically analysed in recent year for a wide range of systems.  
  
In their seminal work \cite{levitovlesovik}, Levitov and Lesovik  
derived the exact formula for the FCS for the electron tunneling  
set--up (single channel):  
\begin{eqnarray}                            \label{LLformula}  
 &~&\ln \chi_{0}(\lambda;V;\{T(\omega)\}) =  {\cal T} \,   
 \int_{-\infty}^\infty \frac{d \omega}{2 \pi}    
 \ln \Big\{   
1 + T(\omega)\\ &\times&\big[ n_L(1-n_R)   
(e^{i \lambda}-1) +  
 n_R(1-n_L)(e^{-i \lambda}-1) \big] \Big\} \,,\nonumber  
\end{eqnarray}  
where ${\cal T}$ is the waiting time, $\lambda$ is the   
measuring field (we will explain this notation in more detail  
shortly), $n_{R(L)}(\omega)=n_F(\omega\pm V/2)$ are the electron  
filling factors in the (right and left) leads,  
$V$ is the bias voltage,   
and $T(\omega)$ is the single electron  
transmission coefficient. The knowledge of $T(\omega)$ thus  
fully defines the FCS for non-interacting systems.   
  
While the Levitov--Lesovik approach can be relatively easily  
generalised to various multi--terminal (multi--channel) set--ups,  
it is notoriously difficult to include electron--electron   
interactions.  
Up to now most works in this direction  
relied upon various perturbative expansions,  
in the tunnelling amplitude \cite{bagrets,reznikov} or 
in the interaction strength \cite{kindnaza} (for a recent review 
see \cite{obzor}), as well as calculations at zero 
temperature \cite{weisssaleur}.  
Certainly no paradigm for an interacting FCS emerged as of yet.  
Notable exceptions are the contribution \cite{safi} as well as
recent works by Andreev and Mishchenko and    
Kindermann and Trauzettel \cite{AM,KT}, where the exact  
FCS was calculated for the (single--channel) Coulomb Blockade (CB)  
set--up of Matveev and Furusaki \cite{matveev,fmatveev}. 
We shall establish the precise connection of these works to  
our results. 
  
The purpose of this Letter is to contribute to our understanding 
of interacting FCSs by means of obtaining 
the exact FCS for two particular experimentally relevant  
set--ups: the Kondo dot and the resonant tunneling (RT) between  
two $g=1/2$ Luttinger liquids (LL).  
  
We start with a brief description of the method. Our goal is the
calculation of the generating function $\chi(\lambda) = \sum e^{i  
q \lambda} P_q$ for the probabilities $P_q$ of $q$ electrons being   
transmitted through the system over time ${\cal T}$. We first define
operators $T_i$ transferring one electron through  
the system in the direction of the current ($i=R$) and in the reversed
direction ($i=L$). The electron counting operator on the Keldysh contour
$C$ can then be written down in its canonical form as 
\cite{levitovlesovik}
\begin{eqnarray}          \nonumber  
 T_\lambda= e^{i\lambda(t)/2} \, T_R+e^{-i\lambda(t)/2} \, T_L \; ,    
\end{eqnarray}  
where the measuringing field $\lambda(t)$ is explicitly time dependent,
$\lambda(t)=\lambda\theta (t)\theta({\cal T}-t)$ on the forward path   
and $\lambda(t)=-\lambda\theta (t)\theta({\cal T}-t)$  
on the backward path. According to \cite{levitovlesovik,reznikov}, the
generating function is then given by the expectation value
\begin{equation} \label{chi}   
\chi(\lambda)=\left\langle T_{{\rm C}} \, e^{-i\int\limits_{{\rm C}}   
T_\lambda(t)dt} \right\rangle\;.  
\end{equation} 
In order to calculate $\chi(\lambda)$  
we define a more general functional   
$\chi[\lambda_-(t),\lambda_+(t)]$ formally given by the same 
Eq.~(\ref{chi}) but where $\lambda(t)$ is now understood to be 
an arbitrary function on the Keldysh contour, $\pm$ referring to two different
functions on the time and anti-time ordered halves of the contour.   
Next we assume that both $\lambda_\pm(t)$ change slowly in time. Then,   
neglecting switching effects, 
one obtains at large ${\cal T}$    
\begin{equation} \label{doublechi}
\chi[\lambda_-(t),\lambda_+(t)]=\exp \left\{-i\int\limits_0^{{\cal T}}  
{\cal U}[\lambda_-(t),\lambda_+(t)]dt \right\} \, ,  
\end{equation}   
where ${\cal U}(\lambda_-,\lambda_+)$ is the  
{\it adiabatic potential}.   
Once the adiabatic potential is computed, the statistics    
is recovered from  
$\ln \chi(\lambda)=-i{\cal T}{\cal U}(\lambda,-\lambda)$. As $\lambda_\pm$ are
external parameters, performing the derivative of both (\ref{doublechi}) and 
(\ref{chi}) with respect to, say, $\lambda_-$, with help of the
Feynman--Hellmann theorem \cite{FH} we immediately obtain    
\begin{equation}                    \nonumber 
\frac{\partial}{\partial\lambda_-}{\cal U}(\lambda_-,\lambda_+)=   
\left\langle \frac{\partial T_\lambda(t)}{\partial\lambda_-}  
\right\rangle_\lambda\;,   
\end{equation}   
where we use notation   
\[   
\langle A(t) \rangle_\lambda=\frac{1}{\chi(\lambda_-,\lambda_+)}   
\left\langle T_{{\rm C}}\left\{A(t)e^{-i\int\limits_{{\rm C}}   
T_\lambda(t)dt} \right\} \right\rangle \, . 
\]   
This is somewhat more complicated than the   
usual {\it Hamiltonian formalism} for a 
quasi--stationary situation, we'll give further  
technical details in the long version \cite{long}; in particular,  
we have verified that Eq.~(\ref{LLformula}) comes out correctly  
in the non-interacting case.   
   
In order to study the FCS for the Kondo dot we use the   
bosonization and refermionization approach, originally applied 
to this problem by Emery and Kivelson \cite{EK}  
(see also \cite{book}) and   
refined by Schiller and Hershfield (SH), see  
\cite{SH}. The starting point is the two-channel Kondo Hamiltonian   
(we set $\hbar = v_F = e = k_B = 1$ throughout),  
\[  
 H = H_0 + H_J + H_M + H_V \, ,  
\]  
where, with $\psi_{\alpha, \sigma}$ being the electron   
field operators in the R,L channels,  
\begin{eqnarray}\label{Hkondo}  
 H_0 &=& i \sum_{\alpha=R,L} \sum_{\sigma=\uparrow,\downarrow}   
 \int \, dx \,  
 \psi^\dag_{\alpha \sigma}(x) \partial_x \psi_{\alpha \sigma}(x) \, ,  
 \nonumber \\  
 H_J &=& \sum_{\alpha, \beta = R,L} \sum_{\nu=x,y,z}  
 J_\nu^{\alpha \beta}  
 s^\nu_{\alpha \beta} \tau^\nu \, , \nonumber \\  
 H_V &=& (V/2) \sum_\sigma \int \, dx \, ( \psi^\dag_{L \sigma}  
 \psi_{L \sigma}  
 - \psi^\dag_{R \sigma} \psi_{R \sigma}) \, , \nonumber \\  
 H_M &=&   
 -\mu_B g_i H \tau^z =   
 - \Delta \tau^z \, .  
\end{eqnarray}  
Here $\tau^{\nu=x,y,z}$ are the Pauli matrices for the  
impurity spin and ($\alpha,\beta=R,L$; $\sigma=\uparrow,\downarrow$;
$\sigma^\nu_{\sigma \sigma'}$ are the components of the $\nu$th Pauli matrix)
\[ 
s^\nu_{\alpha \beta}=\sum_{\sigma,\sigma'}\, \psi_{\alpha \sigma}^\dag(0) \,
\sigma^\nu_{\sigma \sigma'} \, \psi_{\beta \sigma'}(0) \, ,
\] 
are the electron   
spin densities in (or across) the leads, biased by a finite $V$.  
The last term in Eq.~(\ref{Hkondo}) stands for the magnetic  
field, $\Delta=\mu_B g_i H$.   
Following SH, we assume $J_x^{\alpha \beta} =  
J_y^{\alpha \beta} = J_\perp^{\alpha \beta}$,   
$J_z^{LL} = J_z^{RR} = J_z$ and $J_z^{LR}=J_z^{RL}=0$.   
The only transport process then allowed is the  
spin-flip tunnelling, so that we obtain for the $T_\lambda$ operator  
\begin{eqnarray}  
T_\lambda = \frac{J_\perp^{RL}}{2}\left( \tau^+  
e^{i\lambda(t)/2} \psi_{R  
  \downarrow}^\dag \psi_{L \uparrow} + \tau^-  
  e^{i \lambda(t) /2} \psi_{R  
  \uparrow}^\dag \psi_{L \downarrow} \right.  
 \nonumber \\  
 \left.  
 + \tau^+ e^{-i \lambda(t)/2} \psi_{L  
  \downarrow}^+ \psi_{R \uparrow} + \tau^- e^{-i \lambda(t)/2} \psi_{L  
  \uparrow}^\dag \psi_{R \downarrow} \right) \, .\nonumber  
\end{eqnarray}  
After bosonization, Emery-Kivelson rotation, and refermionization  
(see details in \cite{SH})  
and going over to the Toulouse point $J_z = 2 \pi$, which is the only
approximation we make, one obtains  
\begin{eqnarray}                         \label{2chmajham}  
 H'=H'_0 - i (J_- \, b \, \xi_f + J_+ \, a \, \eta_f) - i   
 \Delta \, a \, b + T_\lambda\;,  
\end{eqnarray}  
where the counting term is given by  
\begin{eqnarray}                \label{TlambdaKondo}  
T_\lambda =  
- i J_\perp b \, \left[ \xi \, \cos (\lambda/2) - \eta \, \sin  
(\lambda/2) \right] \, ,  
\end{eqnarray}  
with $J_\pm = (J_\perp^{LL} \pm J_\perp^{RR})/\sqrt{2\pi a_0}$,  
$J_\perp = J_\perp^{RL}/\sqrt{2 \pi a_0}$ ($a_0$ is the lattice constant of  
the underlying lattice model) and $a$ and $b$ being local Majorana  
operators originating from the impurity spin.  
The fields $\eta_f$ and $\xi_f$ in the spin--flavour sector are   
equilibrium Majorana fields, whereas $\eta$ and $\xi$ in the  
the charge--flavour sector are biased by $V$,  
\begin{eqnarray}                      \label{H0prime}  
 H_0' &=& i \int\, dx \, \Big[ \eta_f(x)   
 \partial_x \eta_f(x) + \xi_f(x) \partial_x \xi_f(x)  \\ \nonumber  
 &+&\eta(x) \partial_x \eta(x) + \xi(x) \partial_x \xi(x)   
 + V \xi(x) \eta(x) \Big] \, .   
\end{eqnarray}   
Using Eqs.~(\ref{2chmajham})-(\ref{H0prime}) one can   
straightforwardly evaluate the adiabatic potential   
${\cal U}(\lambda_-,\lambda_+)$ as the problem has   
become quadratic in the Majorana fields.  
  
Skipping details of the calculation, we report the resulting exact  
formula for the FCS of the Kondo dot, which is the {\it main 
result} of this paper:  
\begin{widetext}  
\begin{eqnarray}                \label{mostimportant}  
&~&\ln\chi(\lambda)={\cal T}  \int\limits_{0}^\infty   
\frac{d \omega}{2 \pi} \ln \left\{1+T_1(\omega)  
\left[n_L(1-n_R)(e^{2i\lambda}-1)  
+n_R(1-n_L)(e^{-2i\lambda}-1)\right]\right.\label{stat2}\\  
&~&\left.+T_2(\omega)\left[[n_F(1-n_R)+n_L(1-n_F)](e^{i\lambda}-1)+  
[n_F(1-n_L)+n_R(1-n_F)(e^{-i\lambda}-1)]\right]\right\} \, ,\nonumber  
\end{eqnarray}  
where now the filling factors are $n_{R,L} = n_F(\omega \pm V)$  
[$n_F(\omega)$ being the conventional Fermi function],   
not to be confused with notation in Eq.~(\ref{LLformula}).  
The `transmission coefficients' are  
\[  
 T_1 = \frac{\Gamma_\perp^2 (\omega^2 + \Gamma_+^2)}{\left[ \omega^2 -  
 \Delta^2  
 - \Gamma_+(\Gamma_\perp + \Gamma_-)\right]^2 +  
 \omega^2 (\Gamma_+ + \Gamma_-  
 + \Gamma_\perp)^2} \, \, ,  
 T_2 = \frac{2 \Gamma_\perp \Gamma_- (\omega^2+\Gamma_+^2) + 2 \Delta^2  
 \Gamma_\perp \Gamma_+}{\left[ \omega^2 - \Delta^2  
 - \Gamma_+(\Gamma_\perp + \Gamma_-)\right]^2 +  
 \omega^2 (\Gamma_+ + \Gamma_-  
 + \Gamma_\perp)^2} \, \, ,   
 \]  
\end{widetext}  
where $\Gamma_i = J_i^2/2$ ($i=\pm,\perp$).  
  
For small voltages and zero temperature the generating function   
turns out to be quite simple,  
\begin{eqnarray}                     \label{T0statbis}  
 \chi(\lambda)=\left[1 + T_e(e^{i\lambda}-1) \right]^{N} \; ,  
\end{eqnarray}  
where $N={\cal T} V/ \pi$ is the number of the incoming particles during the
time interval ${\cal T}$ and $T_e = \sqrt{T_1(0)}$ is the effective
transmission coefficient. We analysed explicitly the behaviour of the system
around the Toulouse point. The trivialisation
(\ref{T0statbis}) turns out to be robust against departure from this special
point in parameter space \cite{long}. 
While (\ref{T0statbis}) is the standard Levitov--Lesovik result for spinful
systems,  
the physical content of Eq.~(\ref{mostimportant}) is more 
interesting, since it cannot be reduced to binomial statistics as
in Eq.~(\ref{T0statbis}) at finite $T$. One can see that 
the charge current is carried   
by \emph{two} different quasi--particles with charges   
$q_1 =2 e$ and $q_2 = e$. 
We identify the corresponding transmission coefficients as $T_{1}$ and $T_2$, 
respectively. It was realised by SH that at least in the case of finite
magnetic field $\Delta\neq 0$ it can become energetically favourable to tunnel
electron pairs through the impurity rather than single electrons
\cite{SH}. The presence of the term containing $2 \lambda$ in
(\ref{mostimportant}) can be interpreted as a signature of this effect.
The full transport coefficient $T_0$ as calculated by SH turns out   
to be a {\it composite} one and it is recovered from $T_{1,2}$   
through a very simple relation: $T_0 = T_1 + T_2/2$.  
From the point of view of the Kondo physics, the case when  
$T_2=0$ and the statistics  
reduces to a modified Levitov--Lesovik formula, 
$ 
\chi(\lambda)=\chi_0^{1/2}(2\lambda,2V,\{T_1(\omega)\}) 
$  
(binomial statistics at $T=0$), 
is the symmetric model in zero field  
(the other, unphysical, case of $T_2=0$ is when $J_\pm=0$). 
We have evaluated the first and the second cumulant  
of the Kondo FCS Eq.~(\ref{mostimportant}) which are the same  
as calculated by SH at all $V$ and $T$ \cite{long}.  
We shall not reproduce these two cumulants here  
and concentrate instead on new results.  
 
The full analytic expression for the third cumulant exists  
but is too lengthy to be given here.  
We shall rather investigate various limits and use numerics  
for the general case. So, at $T=0$ we obtain:   
\begin{eqnarray}\label{ctree}  
\langle \delta q^3\rangle &=& {\cal T} \int_{0}^V  
\frac{d \omega}{2 \pi} \,[T_2+8T_1-3(T_2+2T_1)(T_2+4T_1)  
\nonumber \\ &+&  
2(T_2+2T_1)^3]\; .  
\end{eqnarray}   
In the zero magnetic field it yields the following limiting forms: 
\begin{eqnarray}                \label{T0asympt}  
  \langle \delta q^3\rangle_{V \rightarrow 0} &\approx& {\cal T}  
  \, G_0   
  \frac{2 \Gamma_\perp \Gamma_-(\Gamma_- - \Gamma_\perp)}  
{(\Gamma_\perp + \Gamma_-)^3} \, V \, , \\ \nonumber   
  \langle \delta q^3\rangle_{V \rightarrow \infty} &\approx&  
  {\cal T} \,  
 \pi \, G_0 \, \Gamma_\perp   \, , 
\end{eqnarray}  
where $G_0=1/(2\pi)$ is the conductance quantum. 
At low voltages the cumulant is negative for $\Gamma_- < \Gamma_\perp$.  
Generally, under these conditions the $n$-th cumulant  
appears to possess $n-2$ zeroes as a function of $V$, 
according to numerics. The  
saturation value in the limit $V \rightarrow \infty$ is   
independent of the coupling in the spin--flavour channel because   
the fluctuations in the biased   
conducting charge--flavour channel  
are much more pronounced than those in  
the spin--flavour channel, which experiences only relatively weak   
equilibrium fluctuations.   
  
In the opposite case of near equilibrium all  
odd cumulants $\langle \delta q^{2n+1}\rangle$   
are identically zero,  which can readily be seen from  
Eq.~(\ref{mostimportant}) by substituting $n_{R,L}=n_F$ 
into it. In the limit of low temperatures $T\rightarrow 0$   
we recover the conventional Johnson-Nyquist noise power \cite{SH}.   
Moreover, it can be shown that the leading behaviour in temperature   
of \emph{every} even order  
cumulant in this situation is linear, e.~g. for   
$ \langle \delta q^4\rangle$ we obtain  
\begin{eqnarray}  
  \langle \delta q^4\rangle \approx {\cal T} 4 G_0 T \frac{\Gamma_\perp  
  \Gamma_+ (\Gamma_- \Gamma_+ +  \Delta^2)}{\left[ \Delta^2 +  
  \Gamma_+(\Gamma_\perp + \Gamma_-)\right]^2} \, .   
\end{eqnarray}  
  
\begin{figure}  
\vspace*{0.7cm}  
\psfrag{t}{${\cal T}$} 
\epsfig{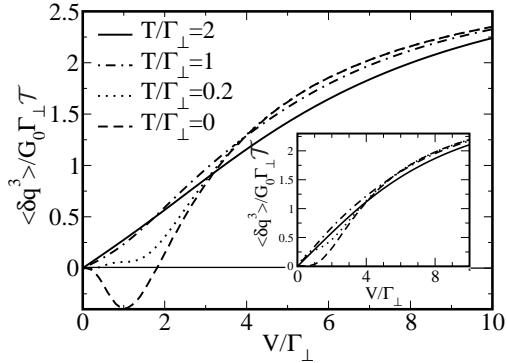} 
\caption[]{\label{C3KondoNew}   
The voltage dependence of the third cumulant for different temperatures  
and zero magnetic field ($\Delta=0$) for   
$\Gamma_-/\Gamma_\perp = 0$ (main graph), and for  
$\Gamma_-/\Gamma_\perp = 0.9$ (inset).}  
\vspace*{0.0cm}  
\end{figure}

For the general situation of arbitrary parameters, the cumulants  
can be calculated numerically.   
The asymptotic value of the third cumulant at high voltages,   
similarly to the findings of \cite{KT}, does not  
depend on temperature and is given by the result (\ref{T0asympt}),   
see Fig.~\ref{C3KondoNew}. In the  
opposite limit of small $V$, $\langle \delta q^3 \rangle$ can be  
negative. Sufficiently large coupling $\Gamma_-$ or magnetic field  
suppress this effect though \cite{long}.   
  
According to the result of Ref.~\cite{reznikov},   
as long as the distribution is binomial,  
$\langle \delta q^3 \rangle/\langle \delta q\rangle=(e^*)^2$,  
where $e^*$ is the effective charge of the current carriers.  
This quantity is to be preferred to the  
Schottky formula because of its weak temperature dependence.   
Indeed we find numerically that the ratio   
$\langle \delta q^3 \rangle/\langle \delta q\rangle$ in the present   
problem is weakly temperature dependent   
(it is flat and levels off to 1) in comparison to   
$\langle \delta q^2 \rangle/\langle \delta q \rangle$, which is indeed in
accordance with (\ref{T0statbis}).  
 
We now briefly turn to the $g=1/2$ RT set--up. This set--up  
has caused much interest recently, see Ref.~\cite{ourPRL} and  
references therein. The Hamiltonian now is   
\begin{equation}\label{12ham}  
H=H_0+\gamma(\psi_L d^\dagger +d\psi_R^\dagger+{\rm H.c})+  
\Delta d^\dagger d+H_C\;,  
\end{equation}  
where $H_0$ stands for two biased LLs at $g=1/2$, $d$ is  
the electron operator on the dot, $\gamma$ is the tunneling  
amplitude and $H_C$ is an electrostatic   
interaction we do not write explicitly here (see \cite{ourPRL}).  
Introducing $\lambda$ as standard, and carrying out   
the bosonization--refermionisation analysis, we find the same set  
of equations as for the Kondo dot, Eq.~(\ref{2chmajham})   
and Eq.(\ref{TlambdaKondo}),  
but with $\lambda\to\lambda/2$ and $J_\perp=2\gamma$, $J_\pm=0$,  
when the Kondo statistics simplifies  
to binomial (unphysical case).  
Consequently, the FCS is given by a modification  
of the Levitov--Lesovik formula:  
\begin{equation}\label{12stat}  
\chi_{1/2}(\lambda)=  
\chi_0^{1/2}(\lambda;2V;\{T_\Delta(\omega)\}) \;, 
\end{equation}  
with the effective transmission coefficient  
$T_{\Delta}(\omega)=4\gamma^4\omega^2/ 
[4\gamma^4\omega^2+(\omega^2-\Delta^2)^2]$  
of the RT set-up in the symmetric case \cite{ourPRL} (the contact asymmetry 
is unimportant).  
All the cumulants are thus obtainable from those 
of the non-interacting statistics Eq.~(\ref{LLformula}). 
 
The $\Delta=0$ 
RT set--up is equivalent to  
the model of direct tunneling between two $g=2$ LLs \cite{ourPRB}.  
The latter model is connected by the strong to weak coupling  
($1/g\to g$) {\it duality} argument to  
the $g=1/2$ Kane and Fisher model \cite{kanefisher}, which is, in turn,  
equivalent to the CB set--up studied  
by KT. Therefore their FCS must be related to our Eq.~(\ref{12stat})   
at $\Delta=0$ by means of the transformation: $T_0\to 1-T_0$   
and $V\to V/2$.   
Indeed after some algebraic manipulation with KT's Eq.~(12), we find   
that the FCS for the CB set--up can be re--written as: 
\begin{equation}\label{CBstat}     \nonumber 
\chi_{CB}(\lambda)= 
\chi_0^{1/2}(-\lambda;V;\{1-T_0(\omega)\}) \;. 
\end{equation}

To summarise, we derived the generating function for the  
charge transfer  
statistics for the Kondo dot in the Toulouse limit and   
analysed the third cumulant in detail. At low temperatures the 
transport is accomplished by electrons as well as 
electron pairs in the generic case
whereas at $T=0$ the conventional binomial statistics is restored.   
  
We wish to thank H.~Grabert, R.~Egger, B.~Trauzettel,   
M.~Kindermann and W.~Belzig for inspiring discussions.  
This work was supported by the Landesstiftung Baden-W\"urttemberg  
(Germany)  
and by the EU RTN DIENOW.  
  
\bibliography{fcsprl}  
  
\end{document}